\documentclass[
    ,final            
  ]
  {aipproc}

\layoutstyle{8x11single}
\usepackage{graphicx}
\usepackage{caption}
\usepackage{subcaption}
\usepackage[pass,letterpaper]{geometry}

\begin{document}

\title{Study of lower hybrid wave propagation in ionized gas by Hamiltonian theory}


\author{A. Casolari}{
  address={ENEA guest}
}

\author{A. Cardinali}{
  address={Associazione Euratom-ENEA sulla Fusione, C.P. 65 - I-00044 - Frascati, Rome, Italy}
}

\begin{abstract}
In order to find an approximate solution to the Vlasov-Maxwell equation system describing the lower hybrid wave propagation in magnetic confined plasmas, the use of the WKB method leads to the ray tracing equations. The Hamiltonian character of the ray tracing equations is investigated analytically and numerically in order to deduce the physical properties of the wave propagating without absorption in the confined plasma. The consequences of the Hamiltonian character of the equations on the travelling wave, in particular, on the evolution of the parallel wavenumber along the propagation path have been accounted and the chaotic diffusion of the timeaveraged parallel wave-number towards higher values has been evaluated. Numerical analysis by means of a Runge-Kutta based algorithm implemented in a ray tracing code supplies the analytical considerations. A numerical tool based on the symplectic integration of the ray trajectories has been developed.
\end{abstract}

\maketitle


\section{Introduction}

Electromagnetic waves in the lower hybrid range of frequencies have long been used as an efficient way of generating a non-inductive current in tokamak plasmas. The problem of describing the propagation of lower hybrid waves in fusion plasmas has been studied intensively both theoretically and in the experiments \cite{Paul Bonoli}. Realistic models make use of the Vlasov-Maxwell system of equations, which gives a precise, although non-intuitive, comprehension of the behavior of the system. Simplified models based on fluid equations and the WKB approximation, instead, give an intuitive understanding of the physical system.\\
The use of WKB approximation is valid in the limit in which the scale length of the disomogeneity is much larger than the wavelength, which is well satisfied in the case of lower hybrid waves. This method allows us to transform the system of integro-differential equations of Vlasov-Maxwell into a more simple  non-linear partial differential equation for the wave-phase surface $S(\vec{x})$. The solution of this equation by the method of characteristics shows its Hamiltonian character, so the evolution of the ray trajectories and the wave vector can be studied as an Hamiltonian flow \cite{Wersinger-Ott-Finn}\cite{Bonoli-Ott}\cite{Paul Bonoli}. This method, also called ray-tracing, allows us to reconstruct the surfaces of constant phase by solving the Hamilton equations for the position and the wave vector, representing the couple of conjugate variables of the system.\\
The dispersion relation of the cold plasma in cylindrical coordinates and without the poloidal magnetic field strictly resembles the Hamiltonian of an isotropic 3D harmonic oscillator, so we studied the effect of poloidal magnetic field and toroidal geometry as small perurbations of the original integrable system. To this end we made use of canonical perturbation theory \cite{Giancarlo Benettin} to find an approximate solution to the problem in terms of the canonical variables action-angle \cite{Bizzarro-Ferreira-Nachak}. We also studied the possiblity of the onset of chaos caused by the non-integrability of the perturbed system because of the toroidal effects. The fact that those effects can be considered as perturbations of the integrable system is made possible by the smallness of the inverse aspect ratio $\epsilon$, which can be taken as a perturbative parameter\\
In addiction to this, we performed a numerical integration of the ray equations both with a Runge-Kutta integrator and with a symplectic algorithm \cite{Channell-Scovel}, which preserves the energy of the system and the volume in the phase space. Both the algorithms gave the same results, which means that the Runge-Kutta method is still valid on the integration time we used. The results show that the average value of $n_{\parallel}$ grows from 2 to a final value between 2.3 and 2.4. We chose to consider the time-averaged value of $n_{\parallel}$ for its istantaneus value is not significant for the wave absorption, despite what is claimed in the reference \cite{Paoletti et al}.

\section{Dispersion relation in cylindrical geometry}

The starting point is the cold electrostatic wave equation suitable to describe the propagation of Lower Hybrid waves:
\begin{equation}
\nabla \cdot (\epsilon^{H(cold)} \cdot \nabla \Phi)=0
\label{electrostatic}
\end{equation}
where $\epsilon^{H(cold)}$ is the Hermitian part of the cold dieletric tensor. The weak non-Hermitian part of it reduces to the Landau damping factor:
\begin{equation}
\gamma=\frac{2\sqrt{\pi}\omega_{p\alpha}^2}{v_{th\alpha}^2}\frac{\omega}{k_{\parallel}v_{th\alpha}}\exp{\left[-\frac{\omega^2}{(k_{\parallel}v_{th\alpha})^2}\right]}
\end{equation}
Eq.(\ref{electrostatic}) can be solved by the WKB expansion, assuming the following form for the scalar potential:
\begin{equation}
\Phi(\vec{x})=\Phi_0(\vec{x})e^{i\delta_0^{-1}S(\vec{x})}
\end{equation}
where $\Phi_0(\vec{x})$ is the slowly varying amplitude, $S(\vec{x})$ is the phase which varies on a smaller scale length, and $\delta_0^{-1}=(\omega a)/c\gg1$ is the WKB expansion parameter ($a$ is the plasma minor radius).\\
Applying the form just chosen for the scalar potential to Eq.(\ref{electrostatic}) and using cylindrical coordinates we obtain the following equation:
\begin{equation}
\epsilon_{\perp}\left[\left(\frac{\partial S}{\partial r}\right)^2+\frac{1}{r^2}\left(\frac{\partial S}{\partial \theta}\right)^2\right]+\epsilon_{\parallel}\left(\frac{\partial S}{\partial z}\right)^2=0
\label{H-J equation}
\end{equation}
This equation represents the Hamilton-Jacobi equation for the generating function $S$. The following step is to identify the derivatives of $S(\vec{x})$ respect to the variables with the corresponding wave vectors, so that we can write the following Hamiltonian:
\begin{equation}
H=n_x^2+\delta_0^2\frac{m_{\theta}^2}{x^2}+\hat{\omega}_{pe0}^2n_z^2x^2=E
\end{equation}
where we defined $\hat{\omega}_{pe0}^2=\omega_{pe0}^2/\omega^2$, $E\approx \hat{\omega}_{pe0}^2n_z^2$ and we assumed a parabolic profile for the density. Here we put $\epsilon_{\perp}=1$ for simplicity and $\epsilon_{\parallel}=1-\hat{\omega}_{pe0}^2(1-x^2)$. The subscript $0$ in the expressions for the plasma frequency refers to the central density. Use has been made of the adimensional variable $x=r/a$ and we introduced the wave numbers in order to make the expression adimensional. $E$ is the equivalent of the energy in the mechanical analogue and is the sum of all the constant terms. This Hamiltonian is perfectly integrable because it is one degree of freedom. To proceed further, we have to find the change from the original variables to the action-angle variables, which are necessary to apply the canonical perturbation theory.\\
The change of variables is formally given by
\begin{equation}
J=\frac{1}{2\pi}\oint p dq\hspace{5mm};\hspace{5mm}\phi=\frac{\partial S}{\partial J}=\frac{\partial}{\partial J}\int pdq
\end{equation}
where the path of integration in the first integral is the path followed by the umperturbed system in its orbit. Since the mechanical analogue of the system we're studying is the isotropic oscillator, in order to calculate the action variable we simply need to integrate between the turning point of the radial motion.\\
After several passages, the final expressions for the action-angle variables are:
\begin{equation}
J=\frac{E}{4\sqrt{A}}-\delta_0\frac{m_{\theta}}{2}\hspace{5mm};\hspace{5mm}\phi=\arcsin\left[\frac{2Ax^2-E}{\sqrt{E^2-4A\delta_0^2m_{\theta}^2}}\right]
\label{change-variables}
\end{equation}
where $A=\hat{\omega}_{pe0}^2n_z^2$, while $m_{\theta}$ is a constant of motion. Inverting the expression for $\phi$ and using the expression for $J$, we obtain:
\begin{equation}
x^2=\frac{1}{\sqrt{A}}(2J+\delta_0m_{\theta}+2\sqrt{J^2+\delta_0Jm_{\theta}}\sin\phi)=\chi+\xi\sin\phi
\end{equation}
Here we have defined for simplicity the functions 
\begin{equation}
\chi=\frac{1}{\sqrt{A}}(2J+\delta_0m_{\theta})\hspace{5mm};\hspace{5mm}\xi=\frac{2}{\sqrt{A}}\sqrt{J^2+\delta_0Jm_{\theta}}
\end{equation}
From Eq.(\ref{change-variables}) we can write the Hamiltonian in terms of the canonical variables: 
\begin{equation}
H_0=2\sqrt{A}(2J+\delta_0m_{\theta})=E
\end{equation}

\section{Effect of the poloidal magnetic field}

When considering the effect of the poloidal magnetic field, the Hamiltonian becomes:
\begin{equation}
H= n_x^2+\left[\delta_0\frac{m_{\theta}}{x}-n_z\frac{\eta x}{1+\alpha x^2}\right]^2+[1-\hat{\omega}_{pe0}^2(1-x^2)-\hat{\omega}_{pe1}^2]\left[\delta_0\frac{\eta m_{\theta}}{1+\alpha x^2}+n_z\right]^2
\label{with_poloid}
\end{equation}
where $\eta=a/R_0$ is the cylindrical analogue of the aspect ratio. In Eq.(\ref{with_poloid}) use has been made of the safety factor
\begin{equation}
q(x)=\eta xB_z/B_{\theta}=1+\alpha x^2
\end{equation}
We have also assumed that $\hat{B_z}=B_z/B\approx 1$ and we have chosen a parabolic profile for the safety factor. Expanding the products, Eq.(\ref{with_poloid}) can be rewritten as
\begin{equation}
H=H_0+ H_1
\end{equation}
where $H_0$ is the Hamiltonian previusly considered, analogue to the isotropic oscillator, while $H_1$ is the perturbation. Taking into account just the main terms, $H_1$ takes the simple form
\begin{equation}
H_1=\frac{E+Fx^2}{1+\alpha x^2}
\end{equation}
where $E=-2\delta_0 n_z\eta m_{\theta}(\hat{\omega}_{pe0}^2+\hat{\omega}_{pe1}^2)$ and $F=2\delta_0 n_z\eta m_{\theta}\hat{\omega}_{pe0}^2$.\\
To find out how much this perturbation affects the dynamic of the unperturbed system, we can estimate its effect on the original frequencies as a first order correction. The frequencies of the unperturbed system are given by:
\begin{equation}
\omega_{0r}=\partial H_0/\partial J=4\sqrt{A}\hspace{5mm};\hspace{5mm}\omega_{0\theta}=\partial H_0/\partial m_{\theta}=2\delta_0\sqrt{A}
\end{equation}
so we see that the ratio of unperturbed frequencies is $\omega_{0r}/\omega_{0\theta}=2/\delta_0\approx 100$. This number gives us an idea of which order of resonance we should expect from perturbation theory, once the toroidal geometry effect is introduced.\\
To calculate the first order modification to the frequency we make use of the formula:
\begin{equation}
\omega_{1r}=\partial \left\langle H_1\right\rangle/\partial J\hspace{5mm};\hspace{5mm}\omega_{1\theta}=\partial \left\langle H_1 \right\rangle/\partial m_{\theta}
\end{equation}
where $\left\langle H_1\right\rangle$ is the perturbation averaged over the angle. After writing $H_1$ in terms of the action-angle variables, the average value of the perturbation is
\begin{equation}
\left\langle H_1\right\rangle=\frac{1}{2\pi}\int_0^{2\pi}\frac{E+F(\chi+\xi\sin \Phi)}{1+\alpha(\chi+\xi \sin\Phi)}d\Phi=\frac{F}{\alpha}-\frac{F-\alpha E}{\alpha \sqrt{a^2-b^2}}
\end{equation}
where the quantities $a=1+\alpha \chi$ and $b=\alpha \xi$ have been introduced. After performing the calculation we find out that those corrections are very small when compared to the unpertubed frequencies, so their effect can be neglected. 

\section{Effect of the toroidal geometry}
Next we examine the toroidal effect on the unpeturbed system. The dispersion relation in toroidal geometry takes the form:
\begin{equation}
H=n_x^2+\left[\delta_0\frac{m_{\theta}}{x}-\frac{\eta n_{\phi}}{1+\epsilon x \cos \theta}\frac{\epsilon x}{1+\alpha x^2}\right]^2+[1-\hat{\omega}_{pe0}^2(1-x^2)-\hat{\omega}_{pe1}^2]\left[\delta_0\frac{m_{\theta}\epsilon}{1+\alpha x^2}+\frac{\eta n_{\phi}}{1+\epsilon x \cos \theta}\right]^2
\label{hamilt_tot}
\end{equation}
where $n_{\phi}$ is the toroidal wave number. To simplify this expression we can expand the denominator because of the smallness of the $\epsilon$ parameter:
\begin{equation}
H=n_x^2+\left[\delta_0\frac{m_{\theta}}{x}-\frac{\eta n_{\phi}\epsilon x}{1+\alpha x^2}(1-\epsilon x \cos \theta)\right]^2+[1-\hat{\omega}_{pe0}^2(1-x^2)-\hat{\omega}_{pe1}^2]\left[\delta_0\frac{m_{\theta}\epsilon}{1+\alpha x^2}+\eta n_{\phi}(1-\epsilon x \cos \theta)\right]^2
\end{equation}
Here again we can separate the expression in different parts:
\begin{equation}
H=H_0+H_1+\epsilon H_2
\end{equation}
where $H_0$ is the unperturbed integrable Hamiltonian, $H_1$ is the 1D perturbation caused by the poloidal magnetic field and $H_2$ is the perturbation which takes into account the toroidal effects. This last part takes the following form:
\begin{equation}
H_2=(Dx+Fx^3)\cos \theta
\end{equation}
where we have considered only the major terms, that is $D\approx 2\hat{\omega}_{pe0}^2\delta_0^2\eta n_{\phi}^2$ and $F=-2\delta_0^2\hat{\omega}_{pe0}^2\eta n_{\phi}^2$. Since they're almost equal in absolute value but opposite in sign, the perturbation can be also written as:
\begin{equation}
H_2\approx D'(x-x^3)\cos \theta
\end{equation}
When evaluating this $D'$ coefficient we find out that it isn't that much smaller than the energy of the unperturbed system, so we expect it to play an important role in the dynamic. In particular we expect that the toroidal geometry affects appreciably the trajectories in the phase space of the system. This form of the perturbation is correct to first order in $\epsilon$, which is a good approximation for $\epsilon<<1$. If we wanted to consider higher order effects we should keep more terms in the expansion of the metric coefficient.

\section{Canonical perturbation theory}
To proceed in the canonical perturbation theory we have to find the generating function $S_1$ for the new variables $J_1$, $\phi_1$, $m_{\theta1}$, $\theta_1$, such that the complete Hamiltonian depends uniquely on the actions.\\
The generating function must fullfill the following equation:
\begin{equation}
\omega_0^{\alpha}\frac{\partial S_1}{\partial \phi_{\alpha}}=\left\langle V \right\rangle-V
\label{omologic}
\end{equation}
where $\omega_0^{\alpha}$ are the unperturbed frequencies, $V$ is the perturbation and $\left\langle V \right\rangle$ is its value averaged over the angle variables. To solve this equation let's call $\tilde{V}=V-\left\langle V \right\rangle$ and expand $S_1$ and $\tilde{V}$ in a Fourier series over the angles:
\begin{equation}
S_1=\sum_m S_{1m} e^{im \cdot \phi}\hspace{5mm};\hspace{5mm}\tilde{V}=\sum_m \tilde{V}_{m} e^{im \cdot \phi}
\end{equation}
so that the solution of Eq.(\ref{omologic}) becomes
\begin{equation}
S_1=i\sum_{m\alpha}\frac{V_{m\alpha}}{m_{\alpha}\omega^{\alpha}}e^{im_{\alpha}\phi^{\alpha}}
\end{equation}
and the new variables are given by
\begin{equation}
J_1=J+\epsilon \partial S_1/\partial \phi\hspace{5mm};\hspace{5mm}\phi_1=\phi-\epsilon \partial S_1/\partial J\hspace{5mm};\hspace{5mm}m_{\theta1}=m_{\theta}+\epsilon \partial S_1/\partial \theta\hspace{5mm};\hspace{5mm}\theta_1=\theta-\epsilon \partial S_1/\partial m_{\theta}
\end{equation}
Please do not mistake $m$ in the Fourier expansion and in the expression for $S_1$ with $m_{\theta}$, which is the poloidal wave number. Since $S_1$ presents the denominators $m\cdot \omega_0$, canonical perturbation theory fails in presence of resonances, that is when $m\cdot \omega_0\approx0$. This is the famous problem of small denominators, which is fundamental for KAM theory. Here $m\cdot \omega_0$ is a short notation for $m_1 \omega_{01}+m_2 \omega_{02}$, meaning that it is a scalar product between an integer vector and a vector whose components are the unperturbed frequencies.\\
To apply the perturbation theory to the Hamiltonian $H_2$ we have to calculate first its Fourier coefficients. First of all the constant term is zero beacause of the $\cos \theta$ factor. Furthermore, the expansion has infinite terms due to the particular form of the perturbation, so we should in principle express it as an infinite series. To calculate this expansion we consider that
\begin{equation}
x-x^3=(\chi+\xi \sin\phi)^{1/2}-(\chi+\xi \sin\phi)^{3/2} \approx \chi^{1/2}(1+\sin\phi)^{1/2}-\chi^{3/2}(1+\sin\phi)^{3/2}
\end{equation}
for $\chi\approx \xi$. The functions $(1+\sin\phi)^{1/2}$ and $(1+\sin\phi)^{3/2}$ have an exact expansion in Fourier series
\begin{equation}
(1+\sin\Phi)^{1/2}=4\sqrt{2}\left[1+\frac{1}{3}\sin\Phi+\frac{1}{15}\cos(2\Phi)-\frac{1}{35}\sin (3\Phi)-\frac{1}{63}\cos (4\Phi)+\cdots\right]
\end{equation}
\begin{equation}
(1+\sin \Phi)^{3/2}=16\sqrt{2}\left[\frac{1}{3}+\frac{1}{5}\sin\Phi-\frac{1}{35}\cos(2\Phi)+\frac{1}{315}\sin (3\Phi)+\frac{1}{1155}\cos (4\Phi)+\cdots\right]
\end{equation}
The denominators in the expansions grow respectively as $D_1(n)\approx(2n)^2$ and $D_2(n)\approx (2n)^4$, so the general term in the expansion becomes rapidly small.\\
We saw in the previous sections that the ratio of unperturbed frequencies is $\omega_{0r}/\omega_{0\theta}\approx 100$, so a resonance between them cannot occur, since the Fourier expansion of $\cos \theta$ has only the term $m=1$.\\
Another approach is possible, that is to expand the metric coefficient
\begin{equation}
\frac{1}{1+\epsilon x \cos \theta}=\sum_{n=0}^{\infty}(-1)^n(\epsilon x \cos \theta)^n
\end{equation}
In this way we can recover all the terms in the Fourier expansion relative to the angle $\theta$, but their amplitude decreases as $\epsilon^n$, so the terms which may be of some interest for the eventual resonance are very small. In addiction of this, expanding this expression in power of $\epsilon$ means going beyond the first order in the perturbation theory, so the entire procedure should be modified.

\section{Numerical results}

The analitical considerations have been supplemented with the numerical integration of ray trajectories by means of a Runge-Kutta algorithm. The position and the wave vector evolve according to the Hamilton equations: applying them to the Hamiltonian Eq.(\ref{hamilt_tot}) we obtain the trajectories in the phase-space of the system, in particular those corresponding to the main variables, that is $x$ and $n_x$. As we could expect, they're not closed due to the non-integrability of the system (figure \ref{fig:phase_space}). In fact the point of closest approach to $x=0$ varies in time due to the variation of $m_{\theta}$.\\
From the numerical solution of Hamiltonian equations it is also possible to study the variation of the parallel wave-number, due to the variation of $m_{\theta}$. The variation we are looking for is an increase in the average value of $n_{\parallel}$, which would be responsible for the wave absorpion by Landau damping on the electrons. The numerical results suggest that $m_{\theta}$ undergoes a small increase, and so does $n_{\parallel}$ (figure \ref{fig:npar_average}).
Even if there is an actual increase in $n_{\parallel}$, it is not enough to justify the wave absorpion, so there must be another mechanism which comes into play in the wave propagation. Another attempt to solve the ray equations by using a symplectic algorithm has been made, and it gave the same results, as a proof of their correctness.\\

\begin{ltxfigure}[tbh]
\begin{minipage}[b]{0.5\linewidth}
  \centering
  \includegraphics[width=0.95\textwidth]{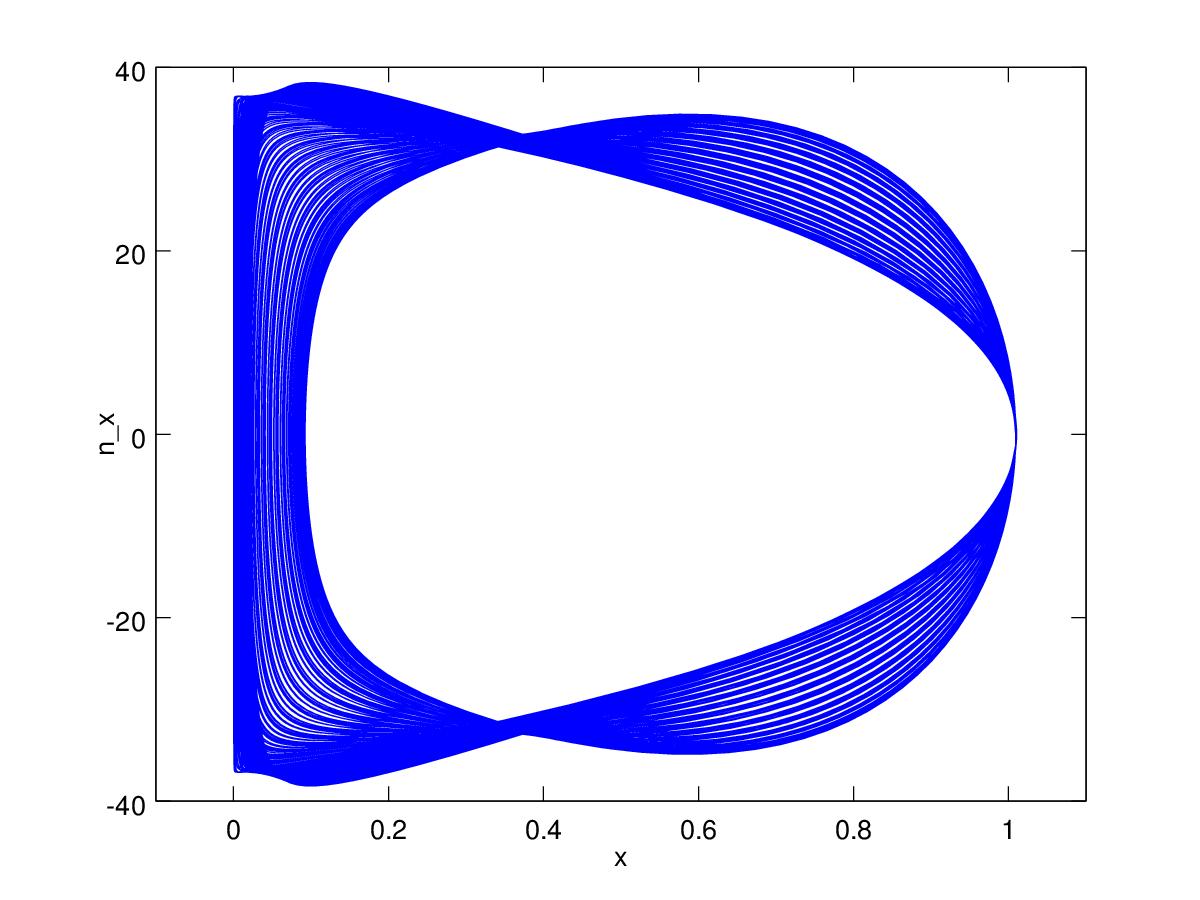}
  \caption{Phase space trajectories}
  \label{fig:phase_space}
\end{minipage}
\hspace{0.2cm}
\hfill
\begin{minipage}[b]{0.5\linewidth}
  \centering
  \includegraphics[width=0.95\textwidth]{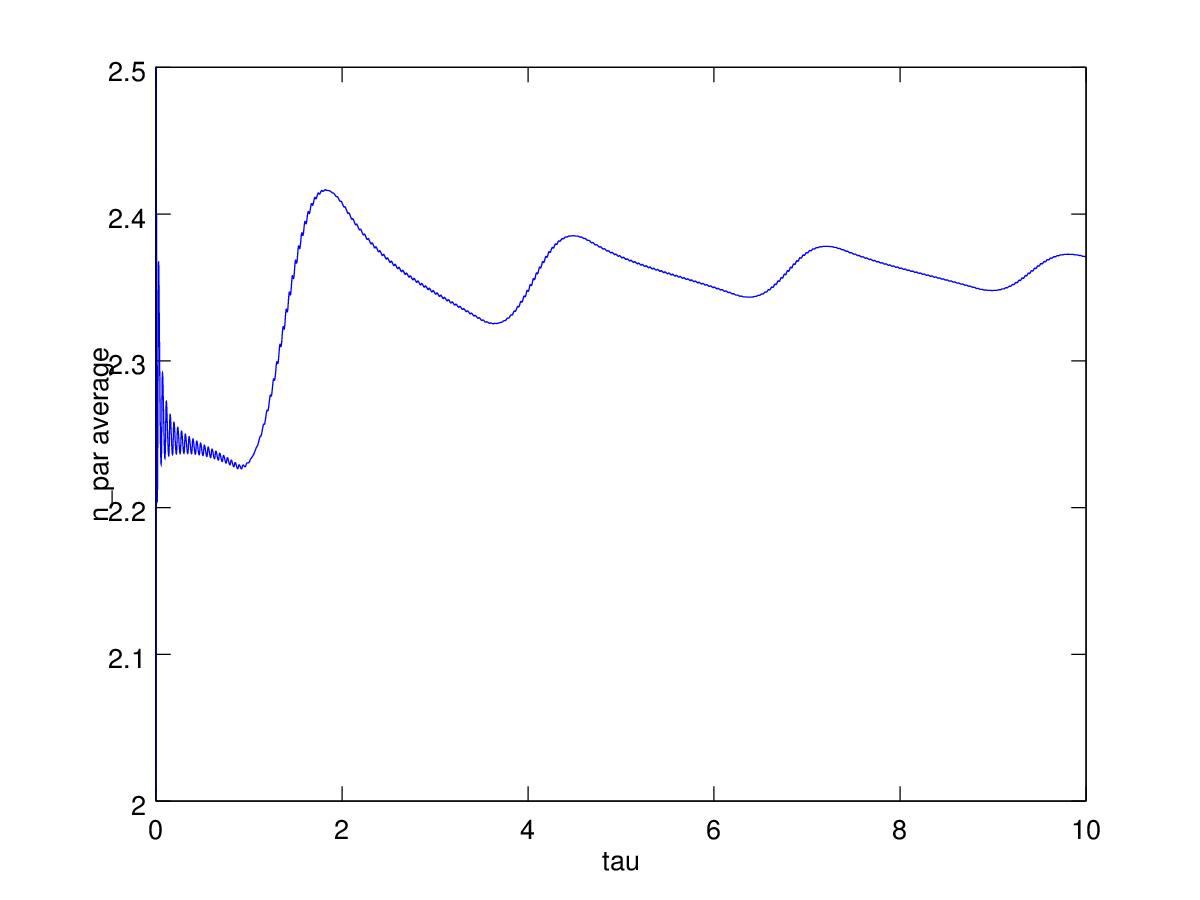}
  \caption{Average value of parallel wave-number}
  \label{fig:npar_average}
\end{minipage}
\end{ltxfigure}

\begin{figure}[htb!]
  \centering
  \includegraphics[width=0.60\textwidth]{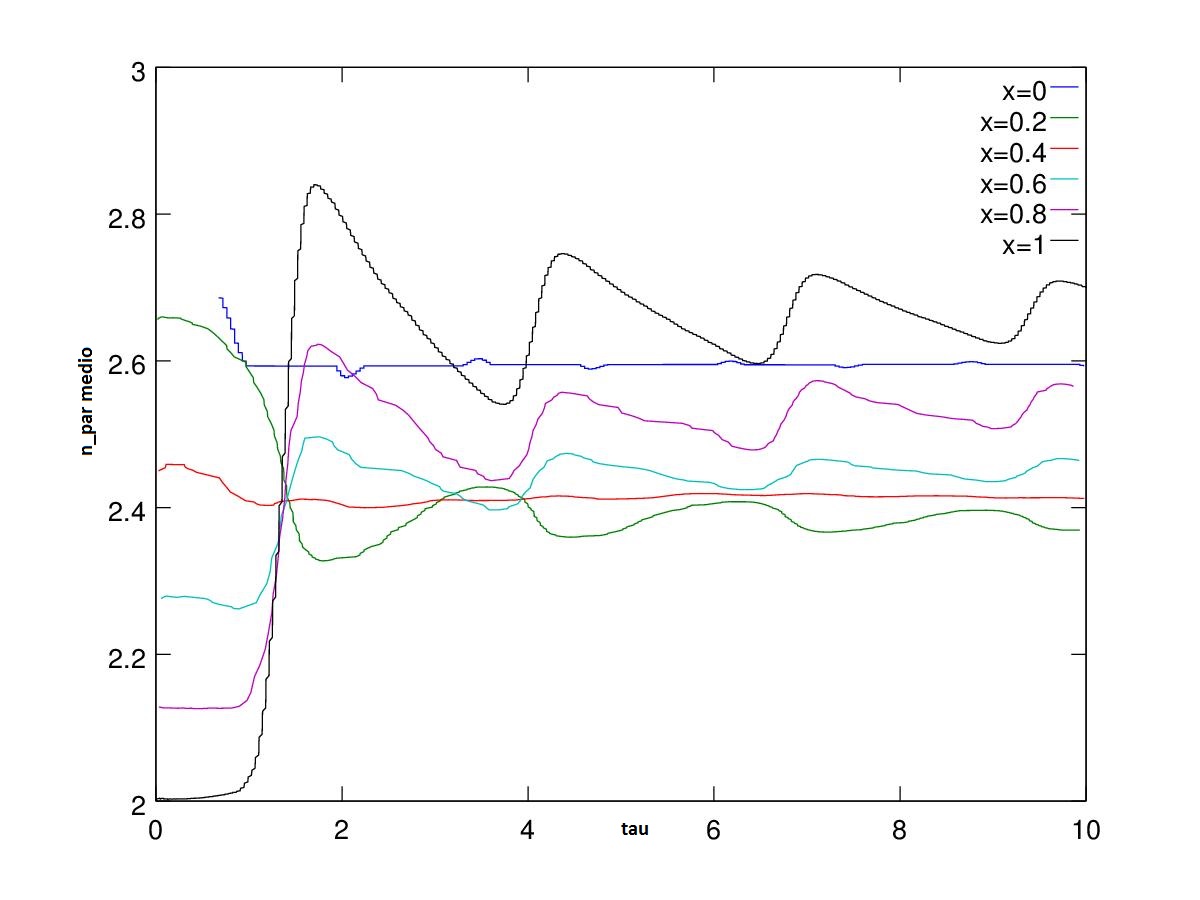}
	\caption{Aritmetic average of $n_{\parallel}$ over different magnetic surfaces}
  \label{fig:medarit}
\end{figure}

We also checked the value of the parallel wave number in different radial positions, corresponding to different magnetic surfaces, in order to see whether the average value of $n_{\parallel}$ grows to the same value in all the volume of the plasma or not. To this end we took the value of $n_{\parallel}$ for different values of the radial coordinate and then we computed the aritmetic average. The results, showed in figure \ref{fig:medarit} indicate that the parallel wave number grows to a value contained between 2.4 and 2.7 over all the magnetic surfaces, but it is a little higher on the boundary and near the magnetic axis.

The results obtained with the Runge-Kutta code and with the symplectic algorithm agree almost perfectly, so there isn't any apparent difference between them. To see where they differ we have to look at the energy conservation in the two different approaches. We see in figure \ref{fig:Erk} and \ref{fig:Esym} the difference between the time evolution of the energy in the two different cases: the Runge-kuta algorithm is such that energy undergoes a temporal increase, whether in the symplectic case its value oscillates about the zero value. In the Runge-Kutta integration we used a low level of accuracy in order to emphasize the increase.

\begin{ltxfigure}[tbh]
\begin{minipage}[b]{0.5\linewidth}
  \centering
  \includegraphics[width=0.95\textwidth]{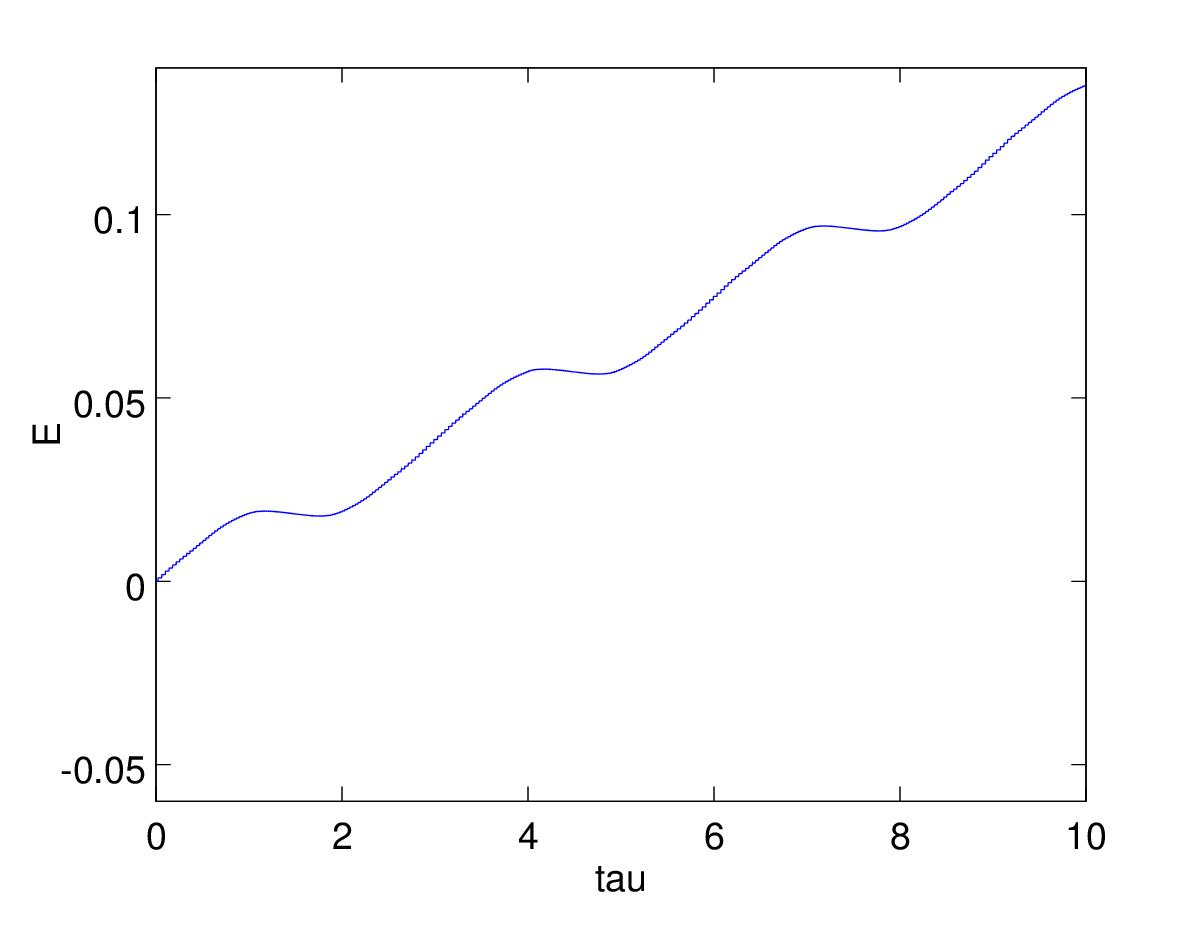}
  \caption{Time evolution of the energy in the Runge-Kutta integration}
  \label{fig:Erk}
\end{minipage}
\hspace{0.2cm}
\hfill
\begin{minipage}[b]{0.5\linewidth}
  \centering
  \includegraphics[width=0.95\textwidth]{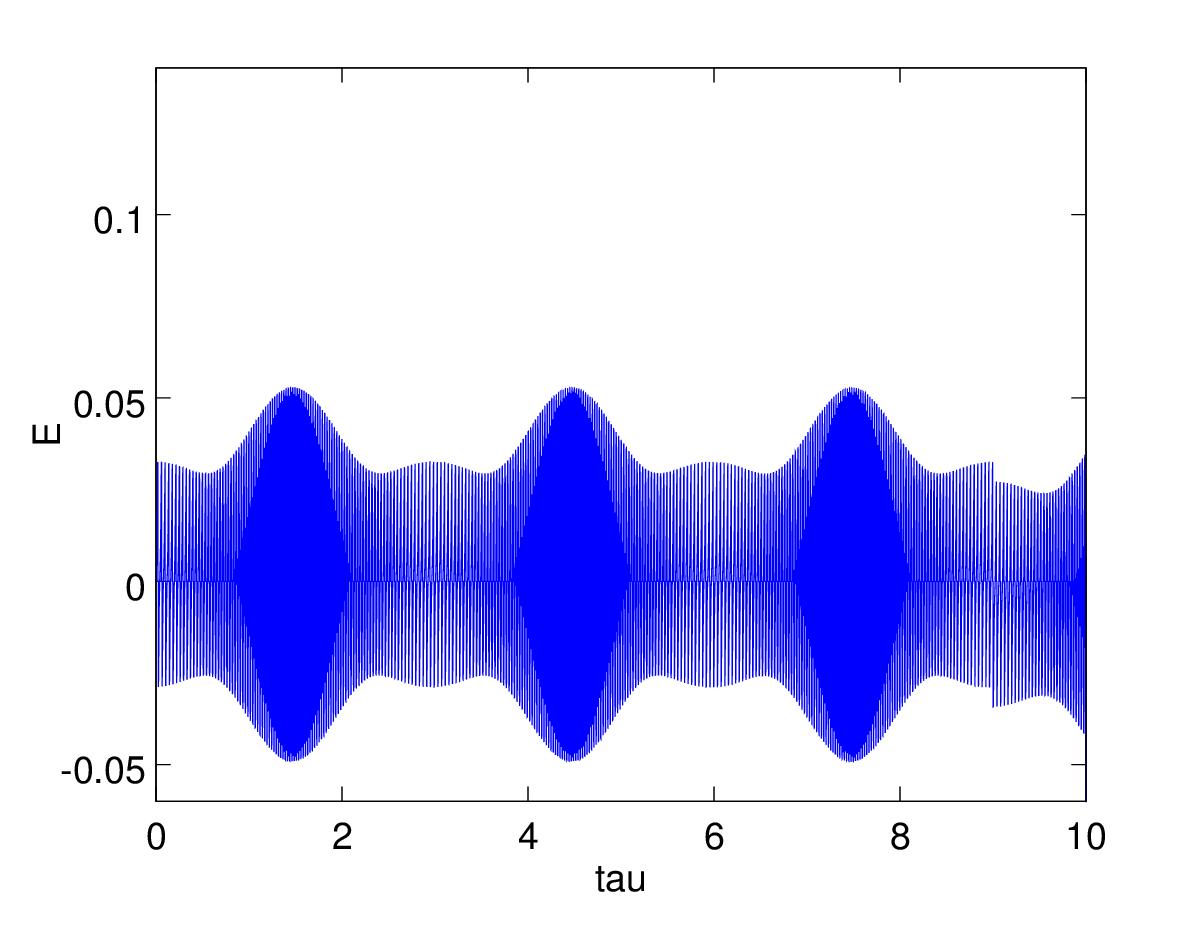}
  \caption{Time evolution of the energy in the symplectic integration}
  \label{fig:Esym}
\end{minipage}
\end{ltxfigure}

We also want to highlight the fact that, for the wave absorpion, we have to consider the average value of $n_{\parallel}$ and not its istantaneus value, for the absorption process takes place on longer time scales than the wave propagation. 
All the considerations we have made are valid in the hypotesis that the magnetic surfaces are circular, for in this case the system shows cylindrical symmetry and the toroidal geometry can be seen as a small perturbation to it. This is the case of several experimental devices such as FTU, Tore Supra and Alcator C.

\appendix
\section{Symplectic algorithm}
Following \cite{Channell-Scovel}, we report the symplectic algorithm, implemented to second order, used to integrate the ray equations.\\
We start with a generating function of the third kind $K=K(p_0,q)$, which generates the variables $p$ and $q_0$ by differentiation:
\begin{equation}
p=-\partial K/\partial q\hspace{5mm};\hspace{5mm}q_0=-\partial K/\partial p_0
\label{equazioni}
\end{equation} 
We expand the generating function in a power series respect to $\tau$, where $\tau$ is the integration step:
\begin{equation}
K(p_0,q)=\sum_{m=0}^{\infty}\frac{\tau^m}{m!}K_m(p_0,q)
\end{equation} 
We define $K_0=-p_0q$ so that the zero order generating function is the identity. In this way we find that
\begin{equation}
p=p_0+\sum_{m=1}^{\infty}\frac{\tau^m}{m!}\frac{\partial K_m(p_0,q)}{\partial q}
\end{equation}
Imposing the condition that
\begin{equation}
\dot{p}=-\partial H/\partial q\hspace{5mm};\hspace{5mm}\dot{q}=\partial H/\partial p
\end{equation}
the generating function becomes:
\begin{equation}
K_0=-p_0q\hspace{2mm};\hspace{2mm}K_1=H\hspace{2mm};\hspace{2mm}K_2=-\frac{\partial H}{\partial q}\frac{\partial H}{\partial p_0}
\label{sistema}
\end{equation}
Once we found these equations, the generating function to the second order is given by :
\begin{equation}
K^{(2)}=K_0+\tau K_1+\frac{\tau^2}{2}K_2
\end{equation}
The Eq.(\ref{equazioni}) reduce to:
\begin{equation}
q=q_0+\sum_{m=1}^2\frac{\tau^m}{m!}\frac{\partial K_m}{\partial p_0}\hspace{2mm};\hspace{2mm}p=p_0-\sum_{m=1}^2\frac{\tau^m}{m!}\frac{\partial K_m}{\partial q}
\label{altre_equazioni}
\end{equation}
The first of the Eq.(\ref{altre_equazioni}) is implicit respect to $q$ and can be solved by imposing $q=q_0$ in the first step. The system of equations must be solved by iteration.


\begin{theacknowledgments}
This work, partially supported by the European Communities under the contract of EURATOM-ENEA-CNR Association, was carried out within the framework of EFDA. The views and opinions expressed herein do not necessarily reflect those of the European Commission.
\end{theacknowledgments}


\bibliographystyle{aipproc}   


\IfFileExists{\jobname.bbl}{}
 {\typeout{}
  \typeout{******************************************}
  \typeout{** Please run "bibtex \jobname" to optain}
  \typeout{** the bibliography and then re-run LaTeX}
  \typeout{** twice to fix the references!}
  \typeout{******************************************}
  \typeout{}
 }

\end{document}